\documentstyle[prl,aps,psfig,twocolumn]{revtex}

\catcode`\@=11

\def\maketitle2{\par 
\begingroup
\let\cite\@bylinecite
\def\thefootnote{\fnsymbol{footnote}}%
\twocolumn[\@maketitle2\vskip2pc]%
\thispagestyle{plain}\@thanks
\endgroup
\def\thefootnote{\arabic{footnote}}%
\setcounter{footnote}{0}%
\let\maketitle2\relax \let\@maketitle2\relax
\let\@thanks\relax \let\@authoraddress\relax \let\@title\relax
\let\@date\relax \let\thanks\relax \let\@abstract\relax 
\let\@pacs\relax}

\def\abstract#1{\gdef\@abstract{{\par 
\bgroup
\ifdim\prevdepth=-1000pt \prevdepth0pt\fi
\hsize\columnwidth
\dimen0=-\prevdepth \advance\dimen0 by17.5pt \nointerlineskip
\small\vrule width 0pt height\dimen0 \relax}{~~}#1\egroup}}

\def\pacs#1{\gdef\@pacs{{\par 
\bgroup
\hsize\columnwidth \parindent0pt
\ifdim\prevdepth=-1000pt \prevdepth0pt\fi
\dimen0=-\prevdepth \advance\dimen0 by20pt\nointerlineskip
\egroup} PACS numbers:~#1}}

\def\@maketitle2{
\@preprint
\@title
\ifdim\prevdepth=-1000pt \prevdepth0pt\fi
\@authoraddress
\@date
\begin{list}{}{\leftmargin=0.10753\textwidth \rightmargin=\leftmargin
\itemsep=1pc\partopsep=-1pc}
\item\@abstract
\item\@pacs
\end{list}
}

\catcode`\@=12

\begin{document}
\draft
\title{Chaos, Order Statistics and Unstable Periodic Orbits}
\author{M.C. Valsakumar, S.V.M. Satyanarayana and S. Kanmani}
\address{Materials Science Division\\
         Indira Gandhi Centre for Atomic Research\\
         Kalpakkam - 603 102\\
         Tamil Nadu, India}
\date{today}

\abstract
{We present a new method for locating unstable periodic points of  
one dimensional chaotic maps. This method is based on
order statistics. The densities of various maxima of the
iterates are 
discontinuous exactly at unstable periodic points of the map. 
This is illustrated using logistic map where densities corresponding to a 
small number of iterates have been obtained in closed form. This technique 
can be applied to the class of continuous time systems where the
successive maxima of 
the time series behave as if they were generated from a unimodal map. 
This is demonstrated using Lorenz model.}

\maketitle2
\narrowtext

Existence of a dense set of unstable periodic orbits (UPO) is 
one of the characteristic properties of a chaotic system \cite{Devaney}.
These orbits represent the skeleton for the strange
attractor of dissipative dynamical systems. Many 
quantities that characterize chaos in the system, such as the fractal
dimension, average
Lyapunov exponent, entropy and the invariant measure of the
corresponding attractor can be determined by knowing the properties of UPO 
\cite{Auerbach,Cvitanovic,Grebogi,Ott,EckRue}. Extraction
of UPO is a necessary step in several studies. For example,   
knowledge of the locations of various cycles is necessary for 
control of chaos\cite{ogy}. Cycles are found to be useful in the
synchronization of chaotic signals \cite{syn}.
Moreover, the
quantum mechanical properties of classically chaotic
conservative systems have, in the semiclassical regime, a
series expansion with respect to the lengths and stability
coefficients of the periodic orbits \cite{Gutzwiller}. The 
importance of unstable periodic orbits and their detection has attracted
the attention of several researchers and a number of numerical 
methods have since been developed to extract unstable periodic orbits 
\cite{Auerbach,Diakonos,So,Biham,Allie,Baba,badii}.

In this letter we present a new technique to extract the UPO of one
dimensional chaotic maps using order statistics. This method works for
all one dimensional chaotic maps for which the existence of invariant density
is guaranteed. 

The theory of extreme values and its generalization to order statistics 
is a classic subject and is extensively used in the study of
independent and identically distributed random variables \cite{lead}. 
Let $\{X_1,X_2, \cdots, X_n\}$ be an
$n$-point set and $M_{n}^{k}\ (1 \le k \le n)$ be its $k^{th}$ maximum
i.e., 
$M_{n}^{n} \le M_{n}^{n-1} \cdots \le M_{n}^{k} \le \cdots M_{n}^{2} \le
M_{n}^{1}$.
Order statistics is the study of distributional properties of
$M_{n}^{k}$. 
In the context one dimensional chaotic map, it has been shown that the
extreme value density is discontinuous on a set of points belonging to 
its unstable periodic orbits \cite{Bala}.
However, this method cannot locate all the periodic points. 
This has motivated us to investigate whether all the unstable periodic 
points can be extracted by using order statistics instead of simple extreme
value analysis. 
 
Let $f(x) : [a,b] \to [a,b]$ be a continuous, one
dimensional chaotic map with an invariant density $\rho_{s}(x)$. Let
$\left\{x_{0},x_{1}=f\left(x_{0}\right),\cdots,
x_{n-1}=f^{n-1}\left(x_{0}\right)\right\}$
be an $n$-point set. The density of the $k^{th}$ maximum is denoted by
$\rho_{n}^{k}(x)$.

Let $S_{n}^{k}$ be the set of locations of the discontinuities of
$\rho_{n}^{k}$. We observe that $O_{n} = \cup_{k=1}^{n} S_{n}^{k}$ for all
$n$, where $O_{n}$ is the set of interior (excluding the end points a,b) 
periodic orbits of all orders
strictly less than $n$. An outline of the proof of the above result is
given towards the end, see for details \cite{pre}. 
This remarkable correspondence between the densities of the
order statistics and UPO enables one to extract the periodic points of any
order by choosing an appropriate $n$. The
locations of the cycles are, thus, read off from the
extremely sharp discontinuities of the densities.

We now illustrate the method by applying it to logistic map, 
$x_{n+1} = f(x_{n}) : [0,1] \to [0,1] = \lambda x_{n}(1-x_{n})$, 
for $\lambda = 4$, 
a well studied unimodal map exhibiting chaos \cite{Devaney}. 

We analytically derive the expression for the densities of the
first and second maxima of a three point set of logistic map using
the formulation described in equation (\ref{prob}). They are 
\begin{equation}\label{lgk1n3}
\rho_{3}^{1} (x) = \cases{{{Sin^{-1}\sqrt{x}} \over {2 \pi}} & 
           $0 \le x < {{3} \over {4}}$\cr
                  {{7 Sin^{-1}\sqrt{x}} \over {2 \pi}} - 1 &
           ${{3} \over {4}} \le x < {{5+\sqrt{5}} \over {8}}$\cr
                  {{6 Sin^{-1}\sqrt{x}} \over {\pi}} - 2 &
           ${{5+\sqrt{5}} \over {8}} \le x <1$\cr}
\end{equation} 
\begin{equation}\label{lgk2n3}
\rho_{3}^{2} (x) = \cases{{{3 Sin^{-1}\sqrt{x}} \over {2 \pi}} &
             $0 \le x < {{5-\sqrt{5}} \over {8}}$\cr
                    {{4 Sin^{-1}\sqrt{x}} \over {\pi}} - {1 \over 2} &
             ${{5-\sqrt{5}} \over {8}} \le x < {3 \over 4}$\cr
                     {{5 Sin^{-1}\sqrt{x}} \over {2 \pi}} & 
              ${{3} \over {4}} \le x < {{5+\sqrt{5}} \over {8}}$\cr
                     {1} &
              ${{5+\sqrt{5}} \over {8}} \le x <1$\cr}
\end{equation} 
Note that the point $x = (5 - \sqrt{5})/8$ is not present in $S_{3}^{1}
= \{3/4,(5 + \sqrt{5})/8\}$, but is a periodic point of order two,
which clearly shows that extreme value statistics is not adequate to extract
all the UPOs and one needs order statistics.

\begin{figure}
\centerline{\psfig{figure=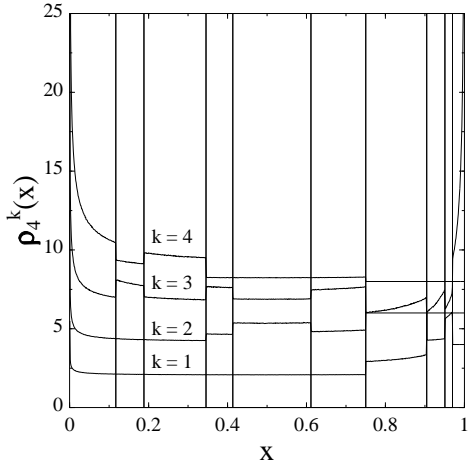,width=7cm,height=7cm}}
\caption{Numerically computed $\rho_{4}^{k}$ for $k=1,2,3\  \rm{and}\  4$ 
of the logistic map ($10^7$ initial conditions, bin width=0.001). 
The locations of the discontinuities and the cycle points of
all orders $ < 4$ are compared. The cycle points are indicated by vertical 
lines. Note the absence of discontinuity at the origin which is a 
fixed point (see the text).}
\label{fig1}
\end{figure}

Numerically, given a map with an invariant density (a mere existence is
necessary and sufficient to render meaning to the averaging procedure), 
one computes 
$\rho_{n}^{k}$ in the following way. Starting with an initial condition,
$n-1$ successive iterates of the map are obtained and arranged in the
ascending order of their magnitude. The $k^{th}$ maximum is thus picked
up. This is repeated for many initial conditions to obtain a histogram
representing $\rho_{n}^{k}$. For the  logistic map $\rho_{4}^{k}(x)$
has been computed numerically and is shown in fig \ref{fig1}. 

Further, the topological entropy, which is a quantitative
characterization of chaos in the system,  can be estimated \cite{balm} 
from the number of periodic orbits as follows,
\begin{equation}\label{topo}
h = {}^{lim}_{n \to \infty}\ \ {1 \over n}\ \ \rm{ln} \ \ N(n)
\end{equation}
where $N(n)$ is the total number of periodic points of order $n$. 
We denote the number of elements in a set $A$ by Card$(A)$. 
Since $N(n)$ =  $ {\rm{Card}}(O_{n+1}) - {\rm{Card}}(O_{n})$, it can be obtained 
from the discontinuities of $\rho_{n+1}^{k}(x) (k=1 \cdots n+1)$ and that of
$\rho_{n}^{k}(x) (k=1 \cdots n)$. In the case of logistic map we obtain
Card$(O_{n})$ as 3, 9, 21, 51 and 105 for $n = $3, 4, 5, 6 and 7 respectively.
Since this formalism is based on the discontinuities of the density over
an interval $[a,b]$, it cannot indicate a periodic point occuring at the extreme
points of the interval $a$ and $b$. Thus, in the above list (see also fig \ref{fig1}), the
fixed point at the origin is not included. The topological entropy calculated 
is 0.6648, which is within $4\%$ of the exact value ln$2$. 

However, for $\lambda < 4$, the 
invariant density, $\rho_{s}(x)$ is not
available in closed form. The existence of $\rho_{s}(x)$ can be checked and 
is sufficient for computing order densities. For most $\lambda < 4$, 
$\rho_{s}(x)$ itself is discontinuous on a large number of points. Thus, 
the corresponding order densities pick up discontinuities at those points 
at which $\rho_{s}(x)$ is discontinuous and also at unstable periodic points. 
Since the strength of the inherent discontinuities of $\rho_{s}(x)$ in $\rho_{n}^{k}(x)$ is large 
compared to that of the discontinuities at periodic points, 
it is numerically difficult to count or detect their locations. 

There are situations where the essence of the dynamics of a continuous
time dynamical system is captured effectively by one dimensional maps or their
equivalent. In his classic
paper Lorenz showed that the successive peaks of a one
dimensional time series behave like iterates of a
map\cite{lorenz}, see also \cite{sparrow}.
We work with the Lorenz model \cite{lorenz} 
\begin{equation}\label{loreq}
\dot{x} = \sigma \left(y - x \right),\hspace{5mm}
\dot{y} = x \left( r - z \right ) - y,\hspace{5mm}
\dot{z} = x y - b z
\end{equation}
for parameters $r=28$, $\sigma = 10$ and  $b=8/3$.

We compute the $\rho_{n}^{k}(x)$ for different $n$ for the map
constructed out of the successive maxima of the time series corresponding 
to the state space variable $z$. Existence of an invariant density for
the so constructed map is verified numerically. 
$\rho_{n}^{k}(x)$ of this map also is discontinuous on a set of points. 
The topological entropy of the
map is  calculated using equation (\ref{topo}). For
n=3, 4, 5 and 6, the topological entropy $h$=0.6486, 0.7361,
0.7700 and 0.7708 respectively, showing a reasonable convergence
in the numerical value of topological entropy.  

It is important to consider the effect of noise on the order densities 
to explore the possible applicability of this method to an experimental 
time series convoluted with noise. 
Here, the distribution 
and strength of the noise play a decisive role. Our preliminary investigations
involve addition of noise, generated from uniform distribution, to the iterates
of the logistic map with $\lambda = 4$. The strength of the noise is $4 \%$ of 
the variance of the iterates. The spill over from [0,1] is reinjected. 
The strength of the discontinuity at the periodic points is observed 
to diminish with increase in the strength of the noise.
   
We now turn to establish the contribution of periodic points to the 
order statistics rigorously.
The formulation of the order statistics involves the probability 
$P_{n}^{k}\left(x\right)$ that
$M_{n}^{k} \in  (x, x+ dx) $ and is given by
\begin{equation}\label{prob}
P_{n}^{k}\left(x\right)  =   
{{1} \over {(k-1)!}} \sum_{i_{1}=0}^{n-1} 
\sum^{}_{{\cal I}} {\rm Prob} \left( \begin{array}{r}
x \le x_{i_{1}} \le x+dx\\
x_{i} > x \vert i \in {\cal I}\\ 
x_{j} < x \vert j \in {\cal J} 
\end{array}
\right)
\end{equation}
where ${\cal I} = \{i_{2},\cdots,i_{k}\}$ and 
${\cal J}=\{i_{k+1},\cdots,i_{n}\}$ with $\{i_{1},\cdots,i_{n}\}$ being 
a permutation of $\{0,\cdots,n-1\}$. Using the joint probability 
density of the $n$ points together with the above equation, we derive
an expression for the cumulative distribution $F_{n}^{k}(x)$ of the 
$k^{th}$ maximum . The order density $\rho_{n}^{k}(x)$ is the derivative 
of $F_{n}^{k}(x)$.

Our central result is : given a continuous one dimensional 
map $f:[a,b] \to [a,b]$ with a continuous invariant density $\rho_{s}(x)$, the 
order density $\rho_{n}^{k}(x)$ is discontinuous at an interior point $x_{\star}$ 
if and only if $x_{\star}$ is an unstable periodic point such that
$f^{l}(x_{\star}) = x_{\star}$ with $l \le n-1$.

We give an outline of the proof and the notation used is described here.
(i) A periodic point of order $l$ is denoted by 
$p_{jl \beta}$, i.e.,  $f^{l}(p_{jl \beta}) = p_{jl \beta}$. 
The index $\beta = 1 \cdots l$ corresponds
to the distinct points of the $l$-cycle. In a chaotic map, in general, there
exist more than one $l$-cycles, and the number of $l$-cycles $N_{l}$ 
increases with $l$. The index $j$ runs over the different $l$-cycles and hence
$j = 1,\cdots,N_{l}$. For example, in  logistic map, we have two orbits of 
period 3. The periodic points of a given $l$-cycle
are ordered such that $p_{jll} = \max_{\hspace{-10pt}{}_{{}_{\beta}}}
\  \{p_{jl\beta}\}$ 
corresponds to the maximum of that cycle.
(ii) Let $y=f^{i}(x)$. The set of preimages of $x$ with respect to $f^{i}$ are
$\left\{ g_{i \alpha}(x) \vert f^{i}\left(g_{i \alpha}\left(x\right)\right)
= x\right\}, \alpha = 1,\cdots, I_{i}$ where $I_{i}$ is the number of 
preimages. For example, the preimages of $x=3/4$  of the logistic
map are $\{3/4,1/4\}$.

As a first step, we show that the extreme value density, 
$\rho_{n}^{1}(x)$ is discontinuous at $x$ if and
only if $x$ is the maximum point of a periodic orbit, i.e, 
$x = p_{j ll }$, $l \le n-1, j = 1, \cdots N_{l}$. Proving this involves two 
stages namely, to show that the extreme value density cannot be discontinuous 
at $x$ unless $x$ is the maximum of a periodic orbit and then show that 
$\rho_{n}^{1}(x)$ is discontinuous at the maximum of all the periodic 
points. The extreme value
density can be formulated by using equation (\ref{prob}) to be
\begin{eqnarray}\label{rhon1}
\nonumber
\rho_{n}^{1}(x)  &=&  \rho_{s}(x) \prod_{j=1}^{n-1} 
 \Theta  \left(x - f^{i}\left( x \right) \right)
 +  \sum_{i=1}^{n-1} \sum_{\alpha=1}^{I_{i}} A_{i\alpha}(x) \\
& \Theta & \left( x - g_{i \alpha}\left( x \right) \right)
\prod_{j \neq i}  \Theta  \left(x - f^{j}\left(g_{i\alpha}\left(x\right)
\right)\right)
\end{eqnarray}
where $A_{i \alpha}(x) = {\rho_{s}\left(g_{i \alpha} \left( x \right) \right)} 
\slash {\vert {{d} \over {dx}} f^{i}(x) \vert_{x = g_{i \alpha}(x)}}$. The above
equation can be recast as 
\begin{equation}\label{recast}
\rho_{n}^{1}(x) = \rho_{s}(x) C_{0}(x) + \sum_{i=1}^{n-1} \sum_{\alpha=1}^{I_{i}}
A_{i\alpha}(x) C_{i \alpha}(x)
\end{equation}

It can be shown that if $\rho_{s}(x)$ is continuous then 
$A_{i \alpha}(x)$ is also continuous, see \cite{pre} for details. 
As all the terms in equation (\ref{rhon1}) are positive
definite, proving either $C_{0}(x)$ or $C_{i \alpha}(x)$ is discontinuous at
a point is sufficient to prove $\rho_{n}^{1} (x)$ to be discontinuous.
Also note $\Theta (z) = (1 + sgn(z))/2$ 
is discontinuous only at $z = 0$.

Since $\rho_s(x)$
and hence $A_{i \alpha}(x)$ are continuous, $\rho_{n}^{1}(x)$ is discontinuous
only if either $\Theta\left(x - f^{l}\left(x\right)\right)$ 
for some $l \le n-1$, or $\Theta \left(x - g_{i \alpha}\left(x\right)\right)$, or
$\Theta \left(x - f^{m} \left(g_{i \alpha} \left(x \right)\right)\right)$ 
for some $m \neq i, m \le n-1$ is discontinuous .

If $\Theta\left(x-f^{l}\left(x\right)\right)$ is discontinuous then 
$x = f^{l}(x)$ implying $x \in \{ p_{j l \beta} \}$. Considering
$C_{0} ( p_{j l \beta} + \epsilon)$ (where $\epsilon$ is an infinitesimal 
quantity of appropriate sign), we have
\begin{equation}\label{pert1}
C_{0}(p_{j l \beta} + \epsilon) = \prod_{k=1}^{n-1} \Theta \left( p_{j l \beta} 
- f^{k}\left(p_{j l \beta} \right) + O (\epsilon) \right)
\end{equation}
If $p_{j l\beta} \neq p_{j ll}$, in the product, there exists a $k_{l}$
such that $f^{k_{l}} (p_{j l \beta}) = p_{jll}$ and the product vanishes 
as $p_{jll} > p_{j l \beta}$ for all $\beta < l$. This shows that $C_{0}(x)$
cannot be discontinuous unless $x = p_{jll}$.

$\Theta \left(x - g_{i \alpha}\left(x\right)\right)$ is discontinuous at
$x = g_{i \alpha}\left(x\right)$. We have
$f^{i}(x) = f^{i}\left(g_{i\alpha}\left(x\right)\right) = x$ and 
this implies $x \in \{p_{ji\beta}\}$.
Considering $C_{i \alpha}(p_{ji\beta} + \epsilon)$, the leading order term is
\begin{eqnarray}\label{cial}
\nonumber
C_{i \alpha}(p_{ji\beta} + \epsilon) =  & \Theta &  \left( \epsilon \left[ 1 - 
{{d} \over {dx}} g_{i \alpha}\left(x\right) \vert_{x=p_{ji \beta}}\right]\right) \\
\prod_{k \neq i} & \Theta & \left( p_{ji\beta} - f^{k} \left(p_{ji\beta}\right)
+ O (\epsilon) \right)
\end{eqnarray}
It can be shown that $g'_{i \alpha}(x) = 1/f'(g_{i \alpha}(x))$ and 
$\vert f'(g_{i \alpha}(p_{ji\beta})) \vert$
is always greater than one for all the unstable periodic orbits. This means
the first $\Theta$ function in the above equation is always non zero. 
Thus, by a similar reasoning outlined above, the second term in the product 
cannot survive unless $x = p_{jii}$.

Consider $\Theta \left( x - f^{m} \left(g_{i \alpha}\left(x\right)\right)\right)$ 
which is discontinuous at $x = f^{m} \left(g_{i\alpha}\left( x\right)\right)$ for some 
$m \in \{1, \cdots,n-1\} \backslash \{i\}$. There exists an $l$ such that
$x = f^{m} \left(g_{i\alpha}\left( x\right)\right)$ implies 
$x = f^{l} \left( x \right)$, where $l = \rm{max} \{m-i,i-m\}$
implying $x \in \{p_{jl\beta}\}$. By using the previous arguments, one can
show that the discontinuity cannot occur unless $x = p_{jll}$.
This proves that $\rho_{n}^{1}(x)$ cannot be discontinuous unless 
$x = p_{jll}$ for some $l \& j$.

Conversely, we show that at every $x=p_{jll}$, 
$\rho_{n}^{1}(x)$ is discontinuous. This is done in two steps for convenience, 
namely (a) if $p_{jll}$ is such that $l > (n-1)/2$, $C_{0}(x)$ 
registers a discontinuity at $x = p_{jll}$ while (b) if $l \le (n-1)/2$, 
$C_{i \alpha}(x)$ will be discontinuous at $x = p_{jll}$. 

(a) Consider $C_{0}(p_{j ll} + \epsilon)$ and writing it as a product of  
three terms, we have
\begin{eqnarray}\label{gt}
\nonumber
C_{0}(p_{j ll} + \epsilon) = 
 & \Theta &  \left( \epsilon \left[1 - {{d} \over {dx}} f^{l}(x)
\vert_{x=p_{jll}}\right]\right)\\
\prod_{k=1}^{l-1} & \Theta & \left(p_{jll} - f^{k}\left(p_{jll}\right) + 
O(\epsilon)\right)\\
\nonumber
\prod_{k=1}^{n-l-1} & \Theta & \left(p_{jll} - f^{k+l}\left(p_{jll}\right) + 
O(\epsilon) \right)
\end{eqnarray}
$f^{k}(p_{jll}) < p_{jll}$ for all $k \le l-1$ and $n-l-1 < l-1$ if 
$l > (n-1)/2$. This implies that the product terms in the equation
(\ref{gt}) are non zero and the first term also can be made non zero 
by choosing an $\epsilon$ of appropriate sign. 
This proves that if $l  > (n-1)/2$, at
$x = p_{jll}, l \le n-1, j = 1,\cdots, N_{l}$, $C_{0}(x)$ and  
hence $\rho_{n}^{1}(x)$ are discontinuous.

(b) We prove that, given $x = p_{jmm}, 
m \neq i$ and $m \le (n-1)/2$, there exists a $\Theta$ function in 
$C_{i \alpha}(x)$, see equation (\ref{rhon1}), with argument 
$x-f^{l}\left(g_{i \alpha}\left(x\right)\right), l \neq i$ such that 
$l = i \pm m$ which is discontinuos at $x=p_{jmm}$, see for 
details \cite{pre} implying $C_{i \alpha}(x)$ and hence $\rho_{n}^{1}(x)$ 
to be discontinuous at $x = p_{jmm}$.

Similarly, one can prove that the density corresponding 
to $k=n$, the minimum of the iterates, $\rho_{n}^{n}(x)$ 
is discontinuous at $x$ if and only if $x = p_{jl1}$, i.e, at the 
minimum of the periodic orbits. 

Finally, it can be proven that at every periodic point $x=p_{jl\beta}$, 
there exists a $k$ such that the order density $\rho_{n}^{k}(x)$ is 
discontinuous at $p_{jl\beta}$. 
This result is obtained by subjecting
$\rho_{n}^{k}(x)$ to a similar analysis outlined in the case of 
$\rho_{n}^{1}(x)$, see \cite{pre} for details.
  
In summary, we have presented a new method to extract all the unstable
periodic points of one dimensional chaotic
maps. This method uses  order statistics. 
We calculate $\rho_{n}^{k}$ 
of the logistic map analytically for $n=3$ and numerically for $n > 3$ 
and illustrate that their discontinuities coincide with the unstable 
periodic points of all orders 
less than $n$.  We demonstrate the applicability of this method to
the map constructed out of
the successive maxima of the time series $z(t)$ of Lorenz equations. In both 
of the above cases topological entropy has been estimated. Further, we give 
an outline of the proof which establishes this connection between order 
statistics and unstable periodic orbits 
for any continuous map with a smooth invariant density.  

We thank K.P.N. Murthy for his careful reading of the manuscript and many 
suggestions. SVMS acknowledges Council of Scientific and Industrial Research, 
India, for Senior Research Fellowship

\end{document}